\documentclass[12pt,english]{article}

\usepackage[T1]{fontenc}
\usepackage[utf8]{inputenc}
\usepackage{geometry}
\usepackage{ulem}
\usepackage{amsmath}
\usepackage{graphicx}
\usepackage{setspace}
\usepackage{esint}
\makeatletter

\providecommand{\tabularnewline}{\\}

\newcommand{\lyxaddress}[1]{
	\par {\raggedright #1
	\vspace{1.4em}
	\noindent\par}
}

\def\be{\begin{equation}}
\def\ee{\end{equation}}

\def\bea{\begin{eqnarray}}
\def\eea{\end{eqnarray}}

\def\ba{\begin{array}}
\def\ea{\end{array}}

\def\bc{\begin{center}}
\def\ec{\end{center}}

\def\bl{\begin{flushleft}}
\def\el{\end{flushleft}}

\def\br{\begin{flushright}}
\def\er{\end{flushright}}

\def\bi{\begin{itemize}}
\def\ei{\end{itemize}}

\def\bt{\begin{tabular}}
\def\et{\end{tabular}}

\numberwithin{equation}{section}

\usepackage[winfonts,UTF8]{ctex}

\usepackage{slashed}
\usepackage{hyperref}
\usepackage{graphicx}
\usepackage[numbers,sort&compress]{natbib}
\date{}

\hypersetup{colorlinks=true,linkcolor=black}

\makeatother

\usepackage{babel}
\begin{document}
\title{Superradiance and stability of the novel 4D charged Einstein-Gauss-Bonnet
black hole}
\author{Cheng-Yong Zhang$^{1}$, Shao-Jun Zhang$^{2,3}$, Peng-Cheng Li$^{4,5}$ ,
Min-Yong Guo$^{4*}$}
\maketitle

\lyxaddress{\begin{center}
\textit{1.Department of Physics and Siyuan Laboratory, Jinan University,
Guangzhou 510632, China}\\
\textit{2.Institute for Theoretical Physics and Cosmology, Zhejiang University of Technology, Hangzhou 310023, China}\\
\textit{3.United Center for Gravitational Wave Physics, Zhejiang University of Technology, Hangzhou 310032, China}\\
\textit{4. Center for High Energy Physics, Peking University, No.5
Yiheyuan Rd, Beijing 100871, P. R. China}\\
\textit{5. Department of Physics and State Key Laboratory of Nuclear
Physics and Technology, Peking University, No.5 Yiheyuan Rd, Beijing
100871, P.R. China}
\par\end{center}}

\begin{abstract}
We investigated the superradiance and stability of the novel 4D charged Einstein-Gauss-Bonnet
black hole which is recently inspired by Glavan and Lin [Phys. Rev. Lett. 124, 081301 (2020)].
We found that the positive Gauss-Bonnet coupling consant $\alpha$ enhances the superradiance, while the negative $\alpha$ suppresses it. The condition for superradiant instability is proved.
We also worked out the quasinormal modes (QNMs) of the charged Einstein-Gauss-Bonnet black hole and found that the real part of  all the QNMs live beyond the superradiance condition and the imaginary parts are all negative. Therefore this black hole is stable. When $\alpha$ makes the black hole extremal, there are normal modes.
\end{abstract}

\vfill{\footnotesize Email: zhangcy@email.jnu.edu.cn,\,\, sjzhang84@hotmail.com, \,\, lipch2019@pku.edu.cn,\,\,   minyongguo@pku.edu.cn. \\  \hspace*{1.5em}  $^*$Corresponding author}

\maketitle
\newpage

\section{Introduction}

As elementary particles, black holes play a central role in gravity including the general relativity and other modified theories of gravity. Numerous studies of past have proven that black holes enjoy many extremely nontrivial effects. One of the most interesting effects is the Penrose progress, which is found to be a new mechanism that energy can be extracted from the Kerr black hole \cite{Penrose:1971uk}. The most essential reason is the existence of ergoregions, where timelike particles can have negative energies. And superradiant effects soon entered people's view as the wave counterpart of the Penross progress \cite{Zeldovic}. Penrose progress and superradiance both require dissipations which can be provided by the ergoregion for an uncharged, stationary and axisymmetric spacetime. But unlike Penrose progress, superradiance can also occur in a nonrotating charged black hole geometry \cite{Bekenstein:1973mi,Hawking:1999dp}, which is fundamentally different from the first situation. Since a spacetime containing a nonrotating charged black hole is believed to be a effectively dissipative environment for charged fields.

Along this line of superradiance, lots of studies have been investigated in many aspects.  On the side of a nonrotating charged black hole, superradiance from RN black holes has been done in \cite{Brito:2015oca} at linearized level by considering a charged scalar field propagating on a RN background with the help of frequency-domain method. The time-domain method was applied to extract amplification factors in \cite{DiMenza:2014vpa}. Also, an analytical treatment was proposed to calculate the amplification factors when the frequency is small \cite{Richartz:2011vf}. These results agree well and support the existence of superradiation in RN black hole geometry mutually. 
Furthermore, some other investigations including superradiance in nonasymptotically flat spacetimes, analogue black hole geometries or higher dimensional spacetimes, superradiance beyond GR  have attracted a lot of attention as well. One is suggested to refer to the comprehensive review  \cite{Brito:2015oca} to overlook the whole picture. Therein, superradiance from black holes in alternative theories of gravity has been studied only in a few cases \cite{Myung:2011we}. Whether superradiance can be stronger in modified theories of gravity still remains a open question \cite{Brito:2015oca}.

As we know, the Einstein-Gauss-Bonnet (EGB) gravity is accounted as one of the most promising candidates for modified gravity \cite{Clifton:2011jh}. And recently, a novel 4D EGB gravity was proposed in \cite{Glavan2019}, where the authors rescale the
Gauss-Bonnet coupling constant $\alpha\to\alpha/(D-4)$ in the limit $D\to4$ and found the corresponding static black hole solution, see also \cite{Cai:2009ua}. This new finding has stimulated a lot of attentions on many fronts
\cite{Konoplya:2020bxa,Guo:2020zmf,Fernandes:2020rpa,Casalino:2020kbt,Wei:2020ght,Kumar:2020owy,
Hegde:2020xlv,Doneva:2020ped,Lu:2020iav,Singh:2020xju, Kobayashi:2020wqy,HosseiniMansoori:2020yfj, Churilova:2020aca,  Mishra:2020gce, Nojiri:2020tph, Liu:2020vkh,SLli:2020,Heydari-Fard:2020,XinghuaJin2020,WenYuanAi2020}.

In this paper, we will explore the superradiance and stability of the novel 4D charged Einstein-Gauss-Bonnet black hole geometry \cite{Fernandes:2020rpa}. Firstly, we determined the constraints on $\alpha$ and the charge of the black hole $Q$ maintaining the event horizon. Then, by considering a charged massless scalar perturbation of the background we obtain the amplification factor of the superradiance and give a detailed analysis of the effects of $\alpha$ under different other parameters. We also pay attention on   stability. We use the asymptotic iteration method (AIM) \cite{Cho2010} to solve the quasinormal modes (QNMs) of the charged scalar perturbation numerically by considering the system as a scattering process. We find there's no instability for the charged 4D EGB black hole under perturbations no matter $\alpha$ is positive or negative in the corresponding allowed region. And very interestingly, we find normal modes are survived for the extremal black hole which is worthy of further studies.

The paper is organized as follows. In section 2, we shortly revisit the novel 4D
EGB gravity and determine the constraints on the GB coupling constant and the charge of the black hole to insure the spacetime contains a static charged black hole. In section 3, we discuss the scalar field
perturbation. And we move to the amplification in section 4. The stabily of the novel 4D charged EGB black hole is investigated in section 5. We summarize our conclusions in section 6.

\section{The spherically symmetric 4D Charged EGB black hole }

The action of the EGB gravity with electromagnetic field in $D$-dimensional
spacetime has the form
\begin{equation}
S=\frac{1}{16\pi}\int d^{D}x\sqrt{-g}\left[R+\frac{\alpha}{D-4}\mathcal{G}^{2}-F_{\mu\nu}F^{\mu\nu}\right].
\end{equation}
Here we have rescaled the coupling constant $\alpha$ by a factor
$\frac{1}{D-4}$. The Gauss-Bonnet term reads
\begin{equation}
\mathcal{G}^{2}=R^{2}-4R_{\mu\nu}R^{\mu\nu}+R_{\mu\nu\alpha\beta}R^{\mu\nu\alpha\beta}=\frac{1}{4}\delta_{\rho\sigma\gamma\delta}^{\mu\nu\alpha\beta}R_{\ \ \mu\nu}^{\rho\sigma}R_{\ \ \alpha\beta}^{\gamma\delta}.
\end{equation}
 The Maxwell tensor $F_{\mu\nu}=\partial_{\mu}A_{\nu}-\partial_{\nu}A_{\mu}$,
in which $A_{\mu}$ is the gauge potential. Varying the action with
respect to the metric, one gets the equation of motion
\begin{equation}
G_{\mu\nu}+\frac{\alpha}{D-4}H_{\mu\nu}=T_{\mu\nu}.\label{eq:EinsteinEq}
\end{equation}
Here $G_{\mu\nu}$ is the Einstein tensor and
\begin{equation}
H_{\mu\nu}=2(RR_{\mu\nu}-2R_{\mu\sigma}R_{\ \nu}^{\sigma}-2R_{\mu\sigma\nu\rho}R^{\sigma\rho}-R_{\mu\sigma\rho\beta}R_{\ \ \ \nu}^{\sigma\rho\beta})-\frac{1}{2}g_{\mu\nu}\mathcal{G}^{2}.
\end{equation}
 The energy-momentum tensor of the Maxwell field takes in this form
\begin{equation}
T_{\mu\nu}=\frac{1}{4}\left(F_{\mu\sigma}F_{\ \nu}^{\sigma}-\frac{1}{4}g_{\mu\nu}F_{\alpha\beta}F^{\alpha\beta}\right).
\end{equation}
 The term $H_{\mu\nu}$ comes from the variation of the Gauss-Bonet
term, which is a topological invariant term in four dimension. Therefore
it does not contribute to the dynamics in four dimension in general.
It can be checked that $H_{\mu\nu}$ always contains a factor $D-4$
and thus disappears when $D=4$. However, by rescaling the coupling
constant $\alpha$, the factor $D-4$ is canceled in (\ref{eq:EinsteinEq}).
Then the Gauss-Bonnet term gives rise to non-trivial dynamics and
novel black hole solutions were discover recently \cite{Glavan2019,Fernandes:2020rpa}.

The spherically symmetric charged black hole solution of (\ref{eq:EinsteinEq})
in four dimension has the form
\begin{equation}
ds^{2}=-f(r)dt^{2}+\frac{1}{f(r)}dr^{2}+r^{2}(d\theta^{2}+\sin^{2}\theta d\phi^{2}),\label{eq:metric}
\end{equation}
 where
\begin{equation}
f(r)=1+\frac{r^{2}}{2\alpha}\left(1\pm\sqrt{1+4\alpha\left(\frac{2M}{r^{3}}-\frac{Q^{2}}{r^{4}}\right)}\right)
\end{equation}
The gauge potential is
\begin{equation}
A=-\frac{Q}{r}dt.
\end{equation}
 Here $M$ is the black hole mass parameter, $Q$ is the charge of
the black hole. In the vanishing limit of $\alpha$ and in the far region,
only the negative branch recovers the Reissner-Nordström (RN) black hole.
Thus we will only study the negative branch in this paper.

\begin{figure}[h]
\begin{centering}
{\footnotesize{}\includegraphics[scale=0.55]{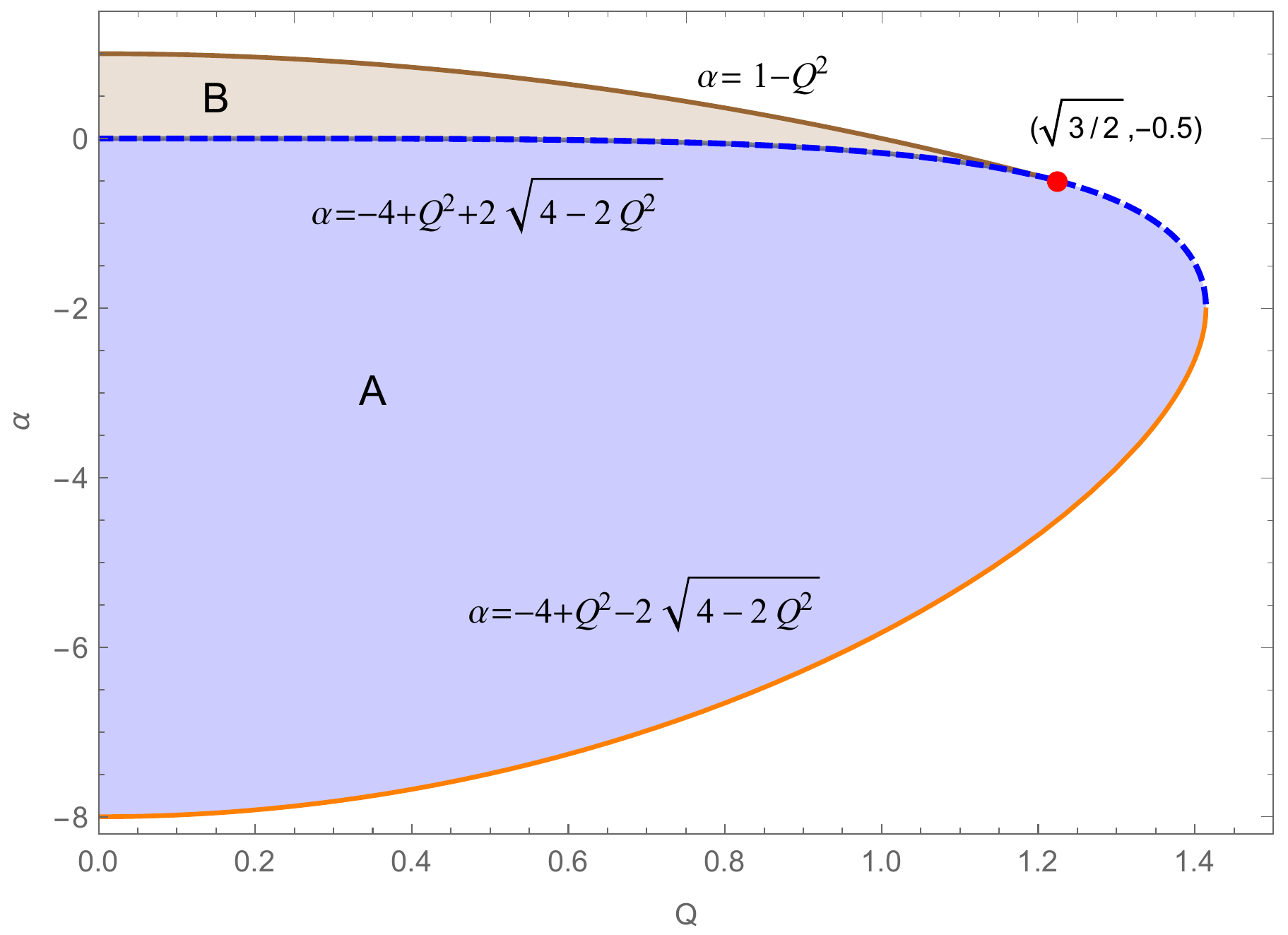}}{\footnotesize\par}
\par\end{centering}
{\footnotesize{}\caption{\label{fig:ParameterRegion} The parameter region where allows the
event horizon $r_{+}$. Here we have fixed $M=1$. Region $B$ corresponds
to $Q^{2}-4-2\sqrt{4-2Q^{2}}<\alpha<1-Q^{2}$ and $0<Q<\sqrt{3/2}$.
Region $A$ corresponds to $Q^{2}-4-2\sqrt{4-2Q^{2}}<\alpha<Q^{2}-4+2\sqrt{4-2Q^{2}}$
and $0<Q<\sqrt{2}$.}
}{\footnotesize\par}
\end{figure}

The solution has  at most two horizons in appropriate parameter region,
\begin{equation}
r_{\pm}=M\pm\sqrt{M^{2}-Q^{2}-\alpha}.\label{eq:horizon}
\end{equation}
 We see that $Q$ can be greater than $M$ when $\alpha$ is negative.
However, $\alpha$ can not be too negative\footnote{Actually when $\alpha$ is negative, the metric function may not be real inside the event horizon. However, since we focus on the region outside the event horizon, we allow $\alpha$ to be negative in this work. See the details in \cite{Guo:2020zmf}.}. It must be ensured that
the metric function is well defined when $r>r_{+}$. We show the allowed
parameter region in Fig. \ref{fig:ParameterRegion}. Hereafter, we
fix $M=1$ for convenience. In region $A$, there is only one horizon
$r_{+}$. In region $B$, there are two horizons $r_{\pm}$. The allowed
region for $\alpha$ is
\begin{equation}
\begin{cases}
Q^{2}-4-2\sqrt{4-2Q^{2}}<\alpha<1-Q^{2}, & \text{when }0<Q<\sqrt{3/2},\\
Q^{2}-4-2\sqrt{4-2Q^{2}}<\alpha<Q^{2}-4+2\sqrt{4-2Q^{2}}, & \text{when }\sqrt{3/2}<Q<\sqrt{2}.
\end{cases}\label{eq:paramRange}
\end{equation}
 Note that when $Q>1$, the solution has no RN
black hole limit since $\alpha$ cannot tend to $0$ now.

{\footnotesize{}}

\section{The charged scalar perturbation}

We consider a charged massless scalar perturbation of the background (\ref{eq:metric}).
It is known that fluctuations of order $\mathcal{O}(\epsilon)$ in
the scalar field in a given background induce changes in the spacetime
geometry of order $\mathcal{O}(\epsilon^{2})$ \cite{Brito:2015oca}.
Therefore to leading order we can study the perturbations on a fixed
background geometry. The massless charged scalar field $\psi$ has
the perturbation equation as
\begin{align}
0= & D^{\mu}D_{\mu}\psi=g^{\mu\nu}\left(\nabla_{\mu}-iqA_{\mu}\right)\left(\nabla_{\nu}-iqA_{\nu}\right)\psi.
\end{align}
 In spherically symmetric background, one can decompose the scalar
field function as
\begin{equation}
\psi=\sum_{lm}\int d\omega e^{-i\omega t}\Psi(r)Y_{lm}(\theta,\phi).
\end{equation}
 Here $Y_{lm}(\theta,\phi)$ is the spherical harmonics on the two
sphere $S^{2}.$ What we are interested in is the radial part of the
equation.
\begin{equation}
0=\frac{1}{r^{2}}\partial_{r}\left(r^{2}f\partial_{r}\Psi\right)+\left(\frac{1}{f}\left(\omega-\frac{qQ}{r}\right)^{2}-\frac{l(l+1)}{r^{2}}\right)\Psi.
\end{equation}
By introducing the tortoise coordinate $r_{\ast}$ and a new variable
$\Psi_{0}$ as
\begin{equation}
dr=fdr_{\ast},\ \Psi=\frac{\Psi_{0}(r)}{r},
\end{equation}
 the radial equation can be written as the Schrödinger-like form
\begin{equation}
0=\frac{\partial^{2}\Psi_{0}}{\partial r_{\ast}^{2}}+\left(\left(\omega-\frac{qQ}{r}\right)^{2}-V_{\text{eff}}\right)\Psi_{0},\label{eq:radEq}
\end{equation}
 where the effective potential reads
\begin{equation}
V_{\text{eff}}=f\left(\frac{l(l+1)}{r^{2}}+\frac{\partial_{r}f}{r}\right).
\end{equation}
 The tortoise coordinate $r_{\ast}$ ranges from $-\infty$ to $\infty$
as $r$ runs from the event horizon $r_{+}$ to infinity. The effective
potential vanishes as $r_{\ast}\to\pm\infty$ and has a potential
barrier in the intermediate region. The asymptotic solution of (\ref{eq:radEq})
can be worked out as
\begin{align}
\Psi_{0}\to & \begin{cases}
\mathcal{T}e^{-i\left(\omega-\frac{qQ}{r_{+}}\right)r_{\ast}}, & r\to r_{+},\\
\mathcal{R}e^{i\omega r_{\ast}}+\mathcal{I}e^{-i\omega r_{\ast}}, & r\to\infty,
\end{cases}\label{eq:AmpBoundary}
\end{align}
 where $\mathcal{I}$ corresponds to the incident amplitude at the
infinity, $\mathcal{T},\mathcal{R}$ are the reflected and transmitted
amplitudes, respectively. Thus we take this problem as a scattering
process. Note that there is no outgoing waves near the event horizon.
Since the background is stationary, the field equation  is invariant
under the transformations $t\to-t$ and $\omega\to-\omega$. There
is another solution which satisfies the complex conjugate boundary
conditions. From (\ref{eq:radEq}), we see that $\Psi_{0}^{\ast}$
(the complex conjugate of $\Psi_{0}$) also satisfy the radial equation.
$\Psi_{0}$ and $\Psi_{0}^{\ast}$ are linearly independent. Their
Wronskian $W=\Psi_{0}\frac{\partial\Psi_{0}^{\ast}}{\partial r_{\ast}}-\Psi_{0}^{\ast}\frac{\partial\Psi_{0}}{\partial r_{\ast}}$
is a constant and independent of $r_{\ast}$. Evaluating $W$ near
the event horizon and at the infinity, we get a relation
\begin{equation}
|\mathcal{R}|^{2}=|\mathcal{I}|^{2}-\frac{1}{\omega}\left(\omega-\frac{qQ}{r_{+}}\right)|\mathcal{T}|^{2}.
\end{equation}
 This relation is independent of the details of the effective potential
barrier. Note that when
\begin{equation}
0<\omega<\frac{qQ}{r_{+}},\label{eq:SuperCondtion}
\end{equation}
 the reflected amplitude $|\mathcal{R}|^{2}$ can be larger than the
incident amplitude $|\mathcal{I}|^{2}$. The wave is amplified. This
phenomenon is called as superradiance. The amplification factor is
defined as

\begin{equation}
Z=|\mathcal{R}|^{2}/|\mathcal{I}|^{2}-1.
\end{equation}
 In the next section, we will study the amplification factor in detail.

\section{The amplification factor of the superradiance}

To work out the amplification factor, one should solve the radial
equation firstly. However, the radial equation is hard to solve analytically
in general. Most of the analytical studies were done with some approximation
such as the frequency tends to zero or the black hole is very small
\cite{Brito:2015oca,Grain2005,Harmark2007,Pappas2016,Ahmed:2016lou,Zhang2017} 
.
In this paper, we solve the
radial equations numerically to study the whole parameter region.
The numerical method adopted here is described in following.

The solution near the event horizon and the infinity can be written
respectively as
\begin{equation}
\Psi_{0}=\begin{cases}
\left(r-r_{+}\right)^{-\frac{i}{2\kappa}\left(\omega-\frac{qQ}{r_{+}}\right)}\sum_{j=0}^{n}a_{i}(r-r_{+})^{j}, & r\to r_{+},\\
e^{i\omega r}\sum_{j=0}^{m}\frac{b_{j}}{r^{j}}+e^{-i\omega r}\sum_{j=0}^{m}\frac{c_{j}}{r^{j}}, & r\to\infty.
\end{cases}\label{eq:asymABC}
\end{equation}
Here $\kappa$ is the surface gravity on the event horizon $r_{+}$.
$n,m$ are the expansion orders. Coefficients $a_{j},b_{j},c_{j}$ depend on frequency
$\omega$. They can be determined by plugging the above expansions into
the radial equation (\ref{eq:radEq}) and comparing the corresponding
coefficients at each order. It can be found that all $a_{j>0}$ are
proportional to $a_{0}$, and $b_{j>0}$ to $b_{0}$, $c_{j>0}$ to
$c_{0}$. We can set $a_{0}=1$ since the radial equation is linear. Given
a frequency $\omega$, the only remaining unknown coefficients are
$b_{0}$ and $c_{0}$. Using the boundary condition near the event
horizon, the radial equation can be integrated numerically outwards.
By comparing the numerical solution with the asymptotic solution (\ref{eq:asymABC})
at infinity, we can get the coefficients $b_{0}$ and $c_{0}$. Note
that $b_{0}$ corresponds to $\mathcal{R}$ and $c_{0}$ to $\mathcal{I}$
in (\ref{eq:AmpBoundary}). Then the amplification factor can be calculated
as $Z=|b_{0}/c_{0}|^{2}-1$.

Since we have fixed $M=1$ in this paper, the free parameters are
$Q,q,\alpha$ and $l$. We analyze their effects on the amplification
factor in detail in this section.

\subsection{The effects of $\alpha$ at different $l$ and $Q$ when $q=1$}

\begin{figure}[h]
\begin{centering}
{\footnotesize{}}%
\begin{tabular}{ccc}
{\footnotesize{}\includegraphics[scale=0.35]{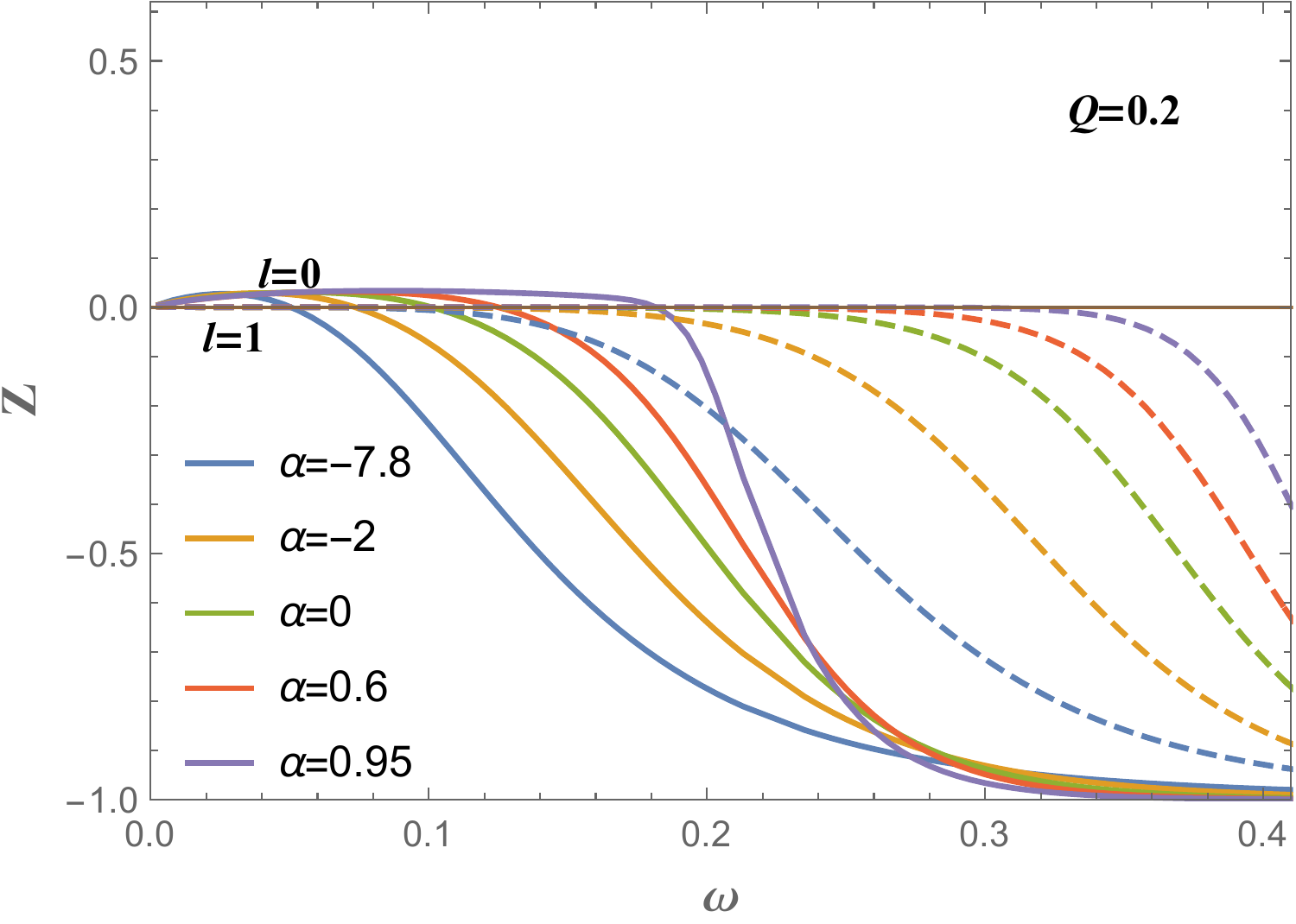}} & {\footnotesize{}\includegraphics[scale=0.35]{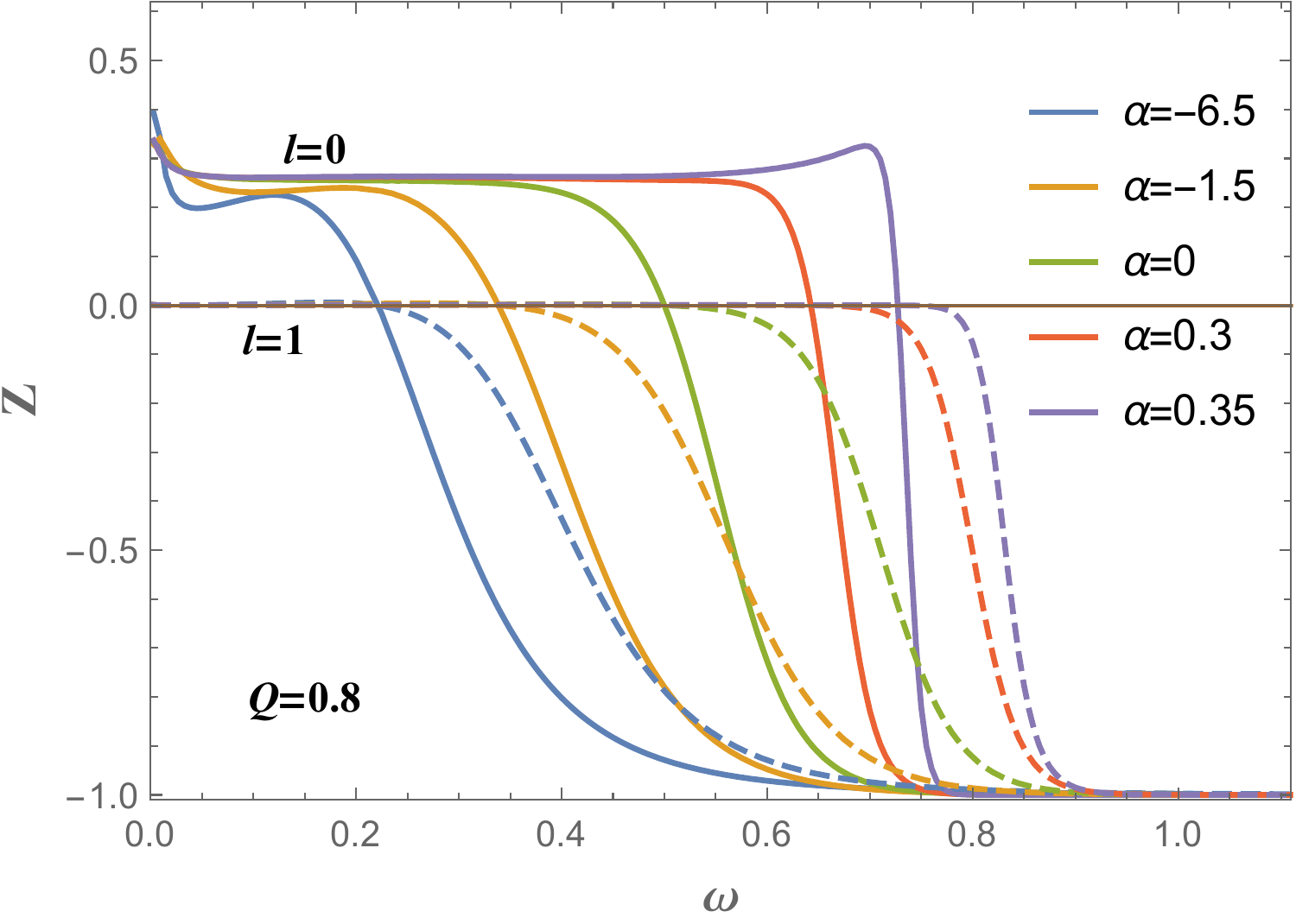}} & {\footnotesize{}\includegraphics[scale=0.35]{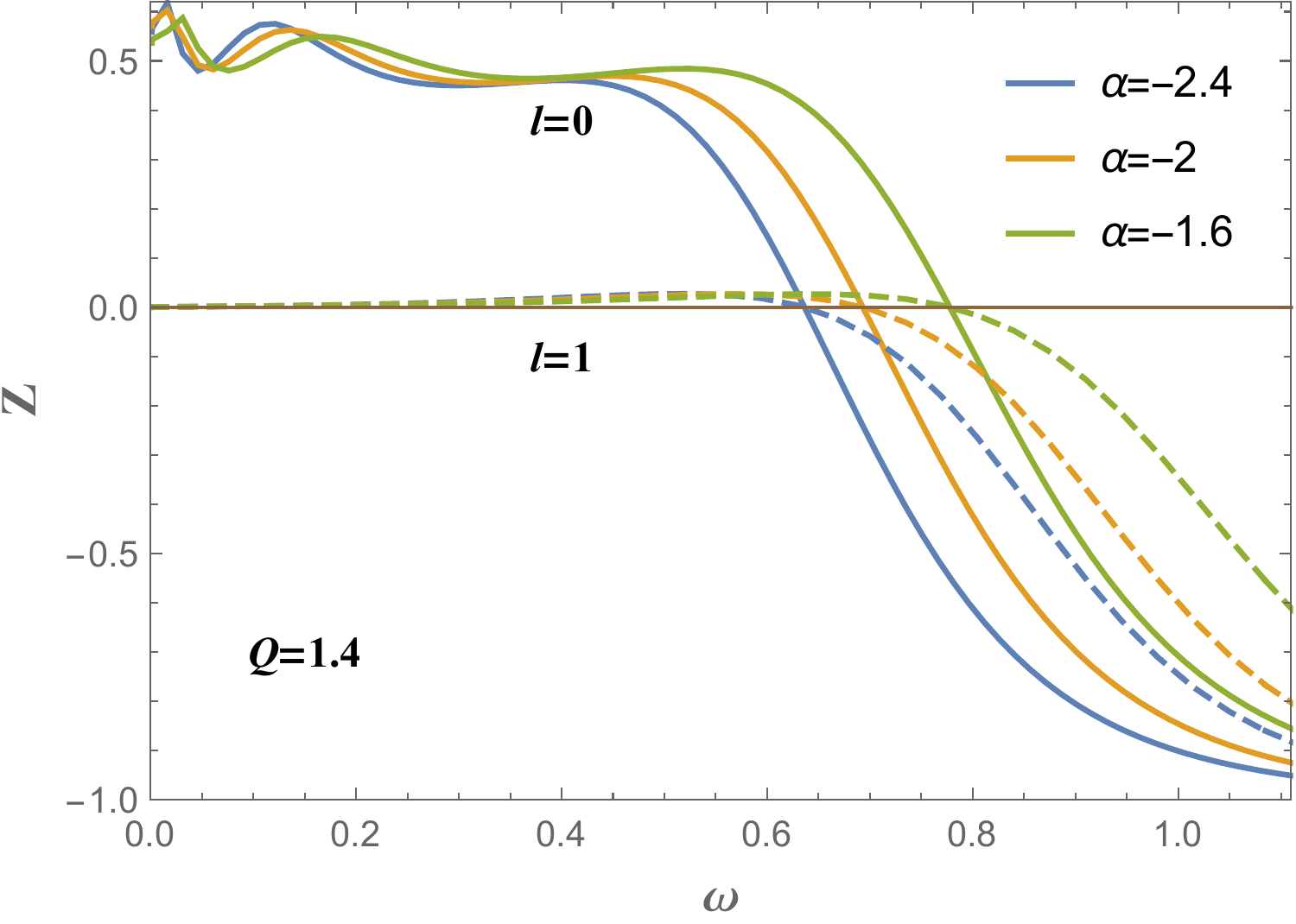}}\tabularnewline
\end{tabular}{\footnotesize\par}
\par\end{centering}
{\footnotesize{}\caption{\label{fig:Ampq1} Effects of $\alpha$ on the amplification factor
when $q=1$. Solid lines for $l=0$, dashed lines for $l=1$. We take
$Q=0.2,0.8$ and $1.4$ as examples to exhibit our results. When $Q=0.2$,
the range of $\alpha$ is $(-7.92,0.96)$. When $Q=0.8$, the range
of $\alpha$ is $(-6.66,0.36)$. When $Q=1.4$, the range of $\alpha$
is $(-2.61,-1.47)$. We vary $\alpha$ in the reasonable region. The
lines with $\alpha=0$ correspond to the RN black hole.}
}{\footnotesize\par}
\end{figure}

The amplification factors when $q=1$ are shown in Fig. \ref{fig:Ampq1}.
Let us consider the case of $l=0$ (the solid lines) first. We see
that the amplification factor is positive just in the region $0<\omega<\frac{qQ}{r_{+}}$.
This implies that it is indeed the superradiance. The superradiance
region is enlarged by positive $\alpha$ and shrunk by negative $\alpha$,
due to the fact that the event horizon $r_{+}$ decreases for positive
$\alpha$ and increases for negative $\alpha$, as can be seen from
(\ref{eq:horizon}). In each panel, we see that the amplification
factor is enhanced by the positive $\alpha$ and suppressed by the
negative $\alpha$. From left panel to right panel, we also see that
the superradiance is enhanced by the black hole charge $Q$. In the
middle panel when $Q=0.8$, there are platforms of the amplification
factor when $\alpha$ tends to make the black hole extremal. The amplification
factor acts as a step-like function of $\omega$. This implies all
modes with frequency satisfying (\ref{eq:SuperCondtion}) are amplified almost equally.
Beyond the superradiant condition, the reflected wave amplitude decreases
so sharply that the amplification factor falls to $-1$.

{\footnotesize{}}

\begin{figure}[h]
\begin{centering}
{\footnotesize{}}%
\begin{tabular}{ccc}
{\footnotesize{}\includegraphics[scale=0.35]{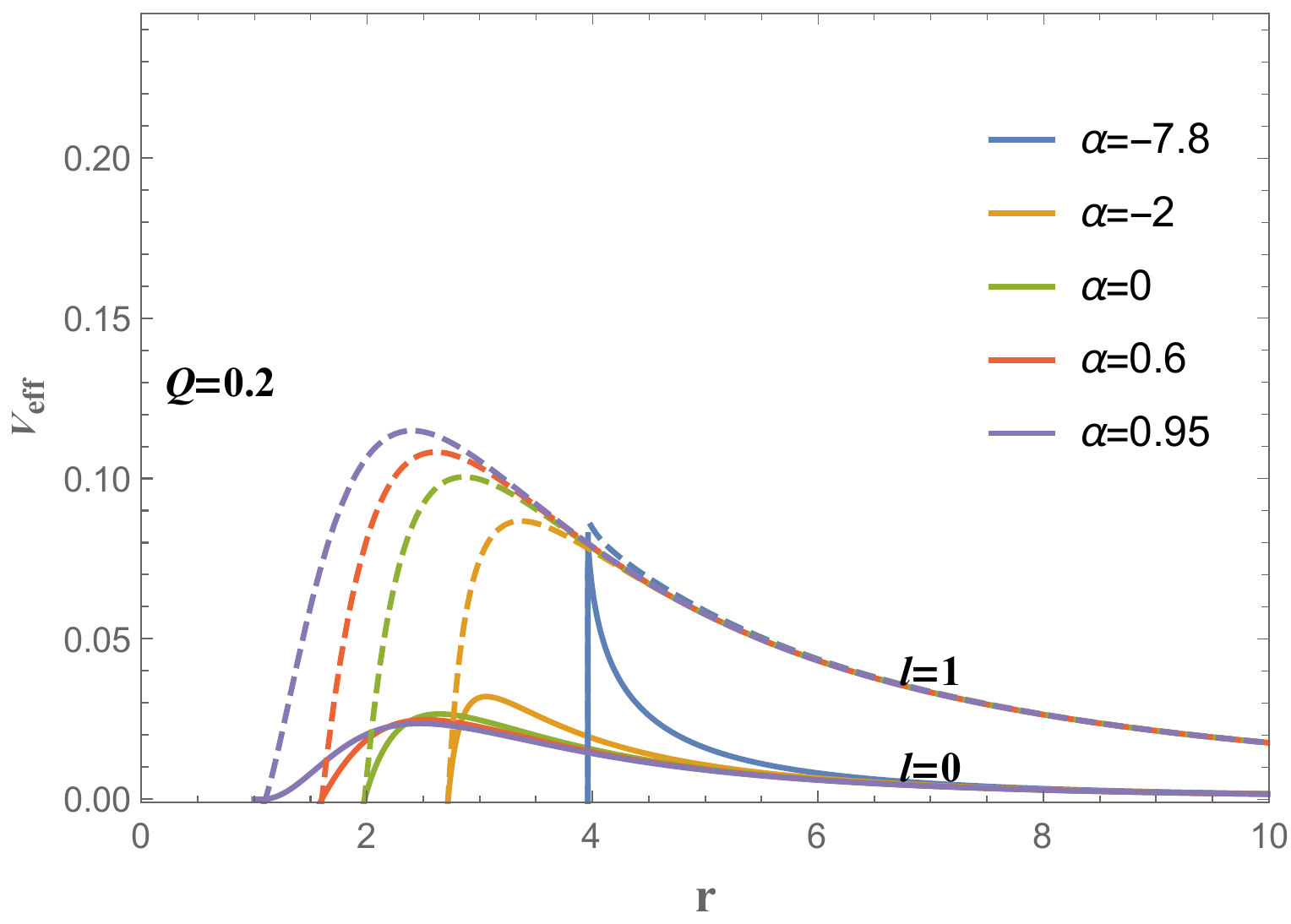}} & {\footnotesize{}\includegraphics[scale=0.35]{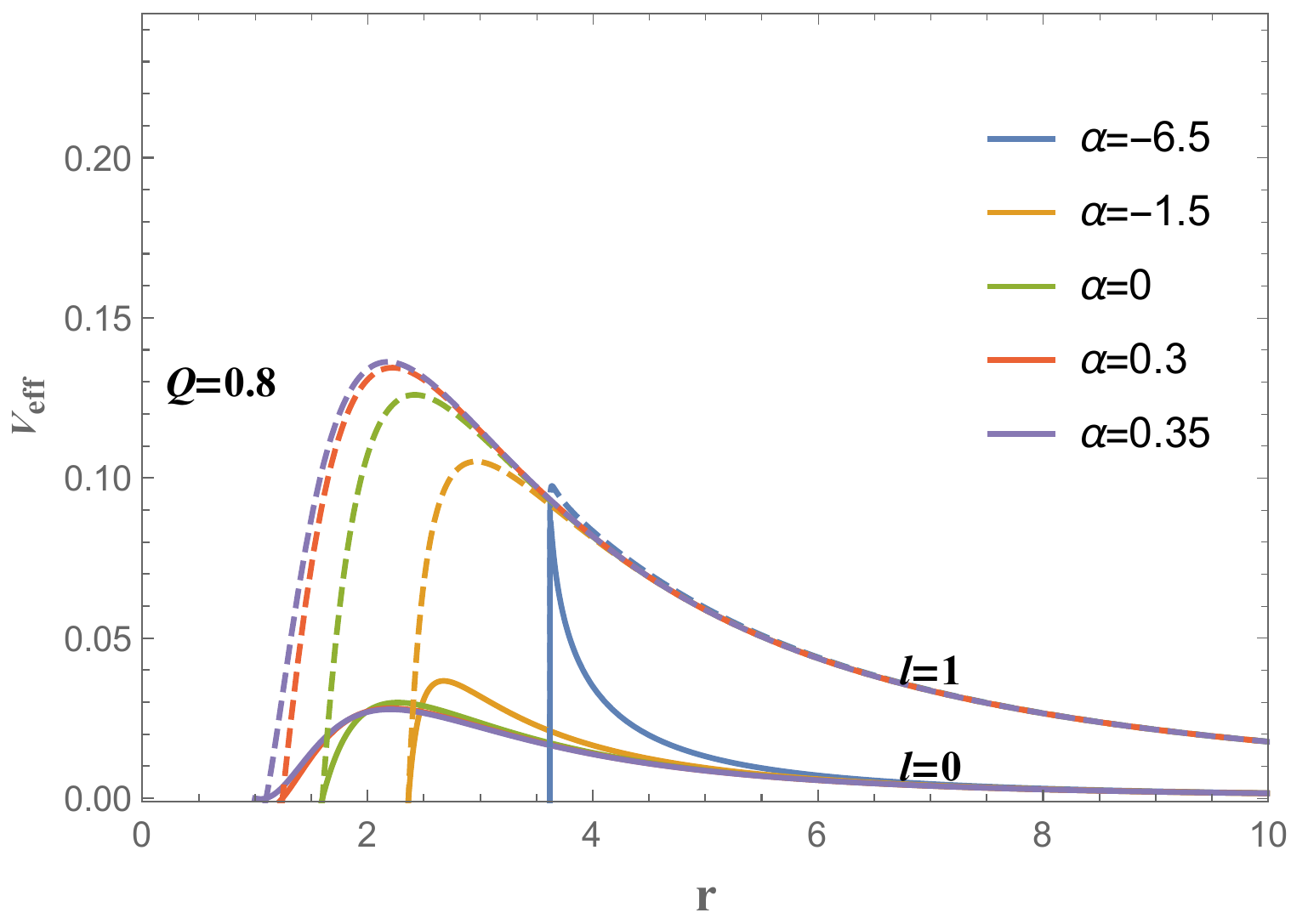}} & {\footnotesize{}\includegraphics[scale=0.35]{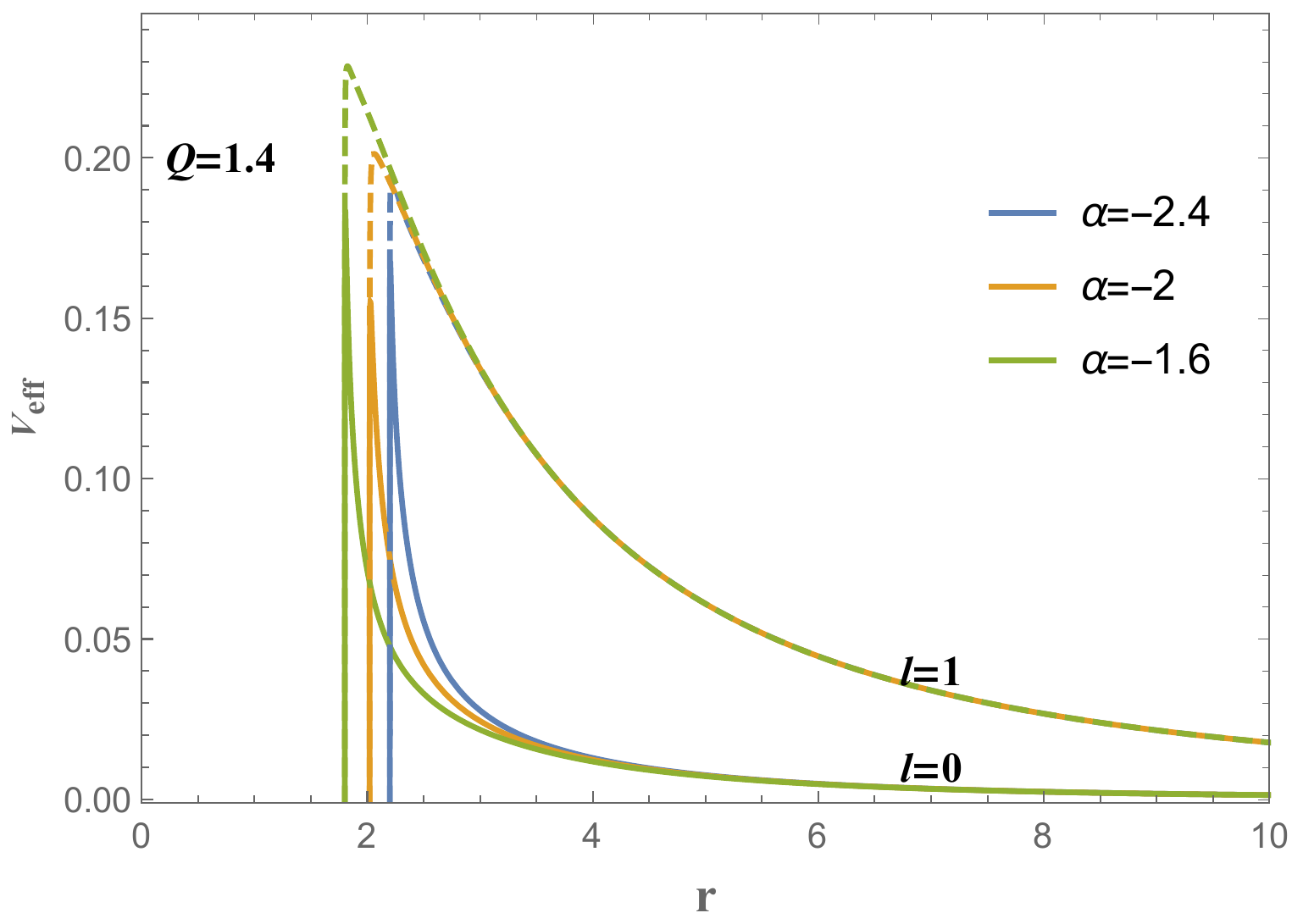}}\tabularnewline
\end{tabular}{\footnotesize\par}
\par\end{centering}
{\footnotesize{}\caption{\label{fig:Ampq1V} The effective potential corresponding to those
with the same parameters in Fig. \ref{fig:Ampq1}. Solid lines for
$l=0$, dashed lines for $l=1$.}
}{\footnotesize\par}
\end{figure}

These  behaviors can be understood intuitively from the effective potential
which were plotted in Fig. \ref{fig:Ampq1V}.
Note that the reflected   amplitude comes from two sides. The one from the extracted energy near the horizon which should cross over the potential barrier to escape to the infinity, we denote as $\mathcal{R}_0$. The one reflected by the effective potential barrier of the incident wave, we denote as $\mathcal{R}_1$.
We see that as $\alpha$
increases in each panel, the effective potential barrier decreases (the solid lines).
The extracted energy from the near horizon can escape to the infinity
easier and leads to a larger reflected amplitude $\mathcal{R}_0$. Therefore the amplification
factor increases as $\alpha$ increases. However, when the frequency
$\omega$ is large enough, the superradiance ceases. Now $\mathcal{R}_0\to 0$ and $\mathcal{R}_1$ dominates. The lower effective
potential barrier, the smaller reflected amplitude $\mathcal{R}_1$. The total amplification
factor should decrease as $\alpha$ increases now.  This phenomenon
can be seen indeed in the far left panel of Fig. \ref{fig:Ampq1}.

{\footnotesize{}}

Now we turn to the cases of $l=1$(the dashed lines in Fig. \ref{fig:Ampq1}
and Fig. \ref{fig:Ampq1V}). In each panel of Fig. \ref{fig:Ampq1},
we see that compared to the cases of $l=0$, the amplification factor is much suppressed in the region
(\ref{eq:SuperCondtion}). But the reflected amplitude is nonzero
due to the superradiance. It is in fact almost equal to the incident
amplitude due to the superradiance. This phenomenon is very different
from the neutral cases, where the reflected amplitude is suppressed
heavily in the whole frequency region for larger $l$. Beyond the
superradiance region (\ref{eq:SuperCondtion}), the amplification
factor of $l=1$ becomes larger than that of $l=0$. This can also
be understood from the effective potential, as shown in Fig. \ref{fig:Ampq1V}.
The effective potential barrier increases with $l$. Beyond the superradiance
region, the waves with lower $l$ are more likely to cross over the potential barrier and be absorbed by the black hole and thus leading to smaller amplification factor. The wave with higher $l$ is more likely to be reflected by the higher barrier. Thus the amplification factor is enhanced for larger $l$.

\subsection{The effects of $\alpha$ at different $l$ and $Q$ when $q=5$}

\begin{figure}[h]
\begin{centering}
{\footnotesize{}}%
\begin{tabular}{ccc}
{\footnotesize{}\includegraphics[scale=0.3]{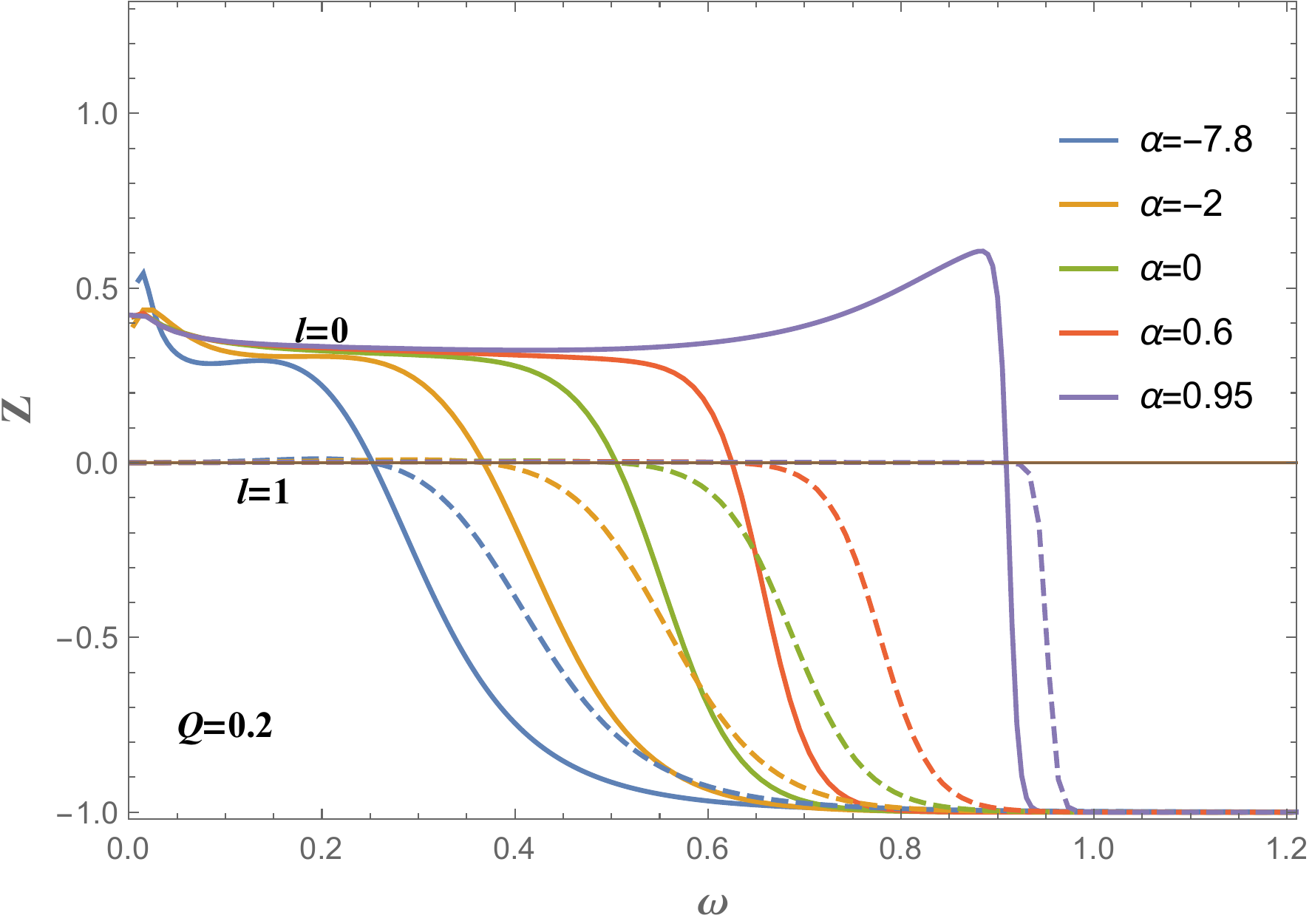}} & {\footnotesize{}\includegraphics[scale=0.3]{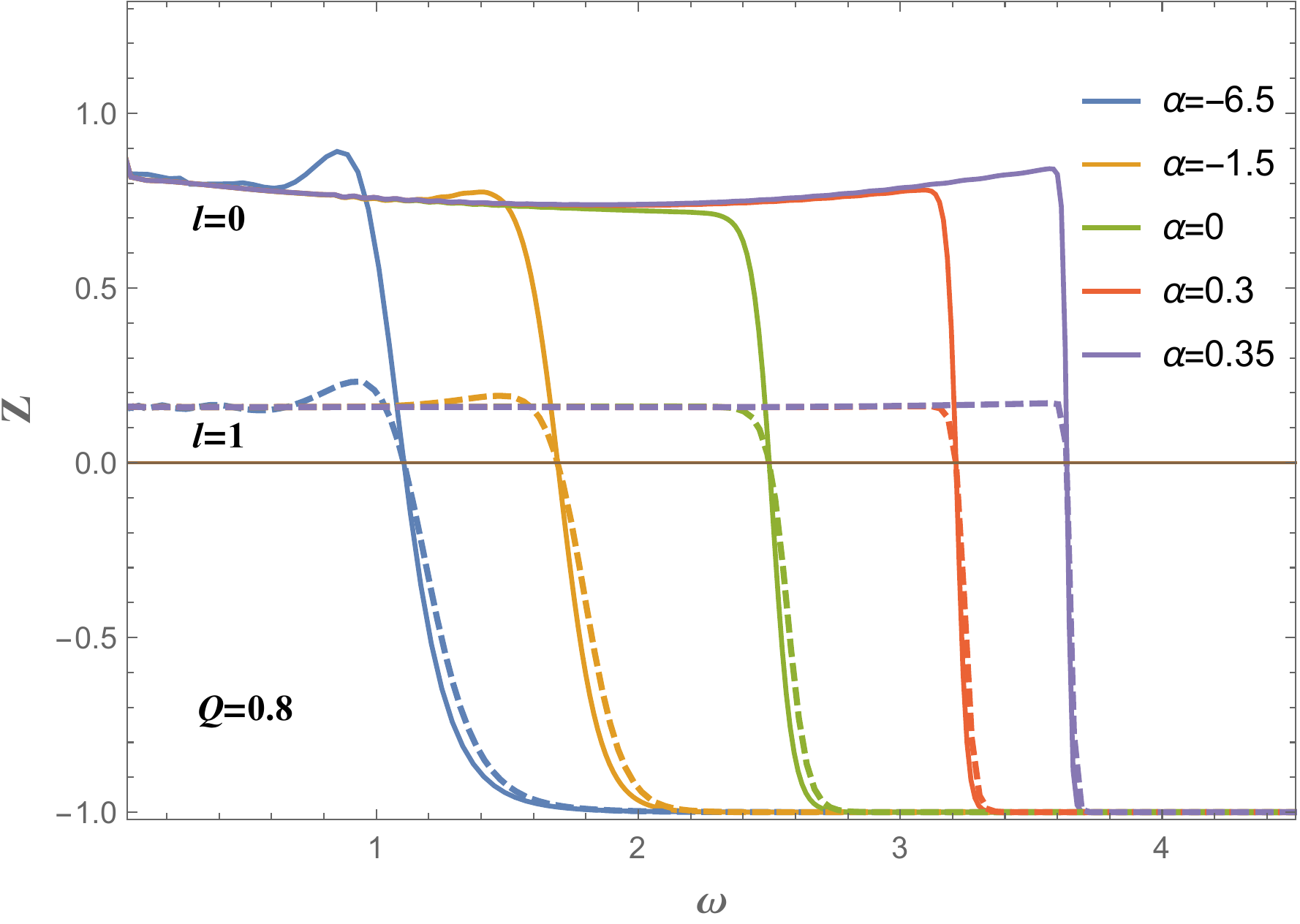}} & {\footnotesize{}\includegraphics[scale=0.3]{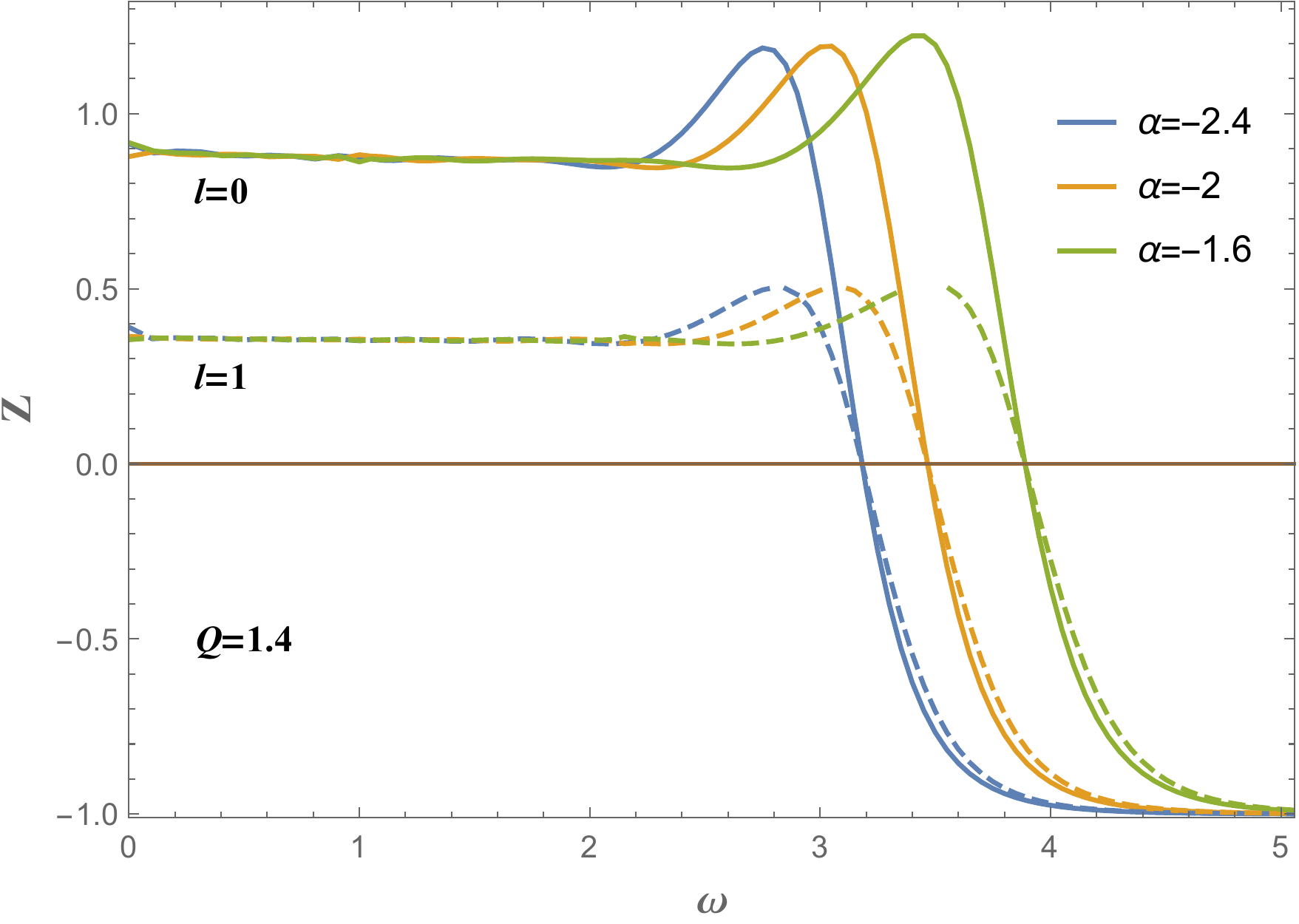}}\tabularnewline
\end{tabular}{\footnotesize\par}
\par\end{centering}
{\footnotesize{}\caption{\label{fig:Ampq5} Effects of $\alpha$ on the amplification factor
$Z$ when $q=5$. The other parameters are the same as those in Fig.
\ref{fig:Ampq1}. The lines with $\alpha=0$ correspond to the cases
of the RN black hole. The small wiggles for small $\omega$ are caused
by numerical error.}
}{\footnotesize\par}
\end{figure}

We take the same parameters as those in the last subsection, except
by setting $q=5$, to study the effect of $q$ on the amplification
factor. Unlike to $Q$, the parameter $q$ is unlimited in principle.
The amplification factors are shown in Fig. \ref{fig:Ampq5}. In each
panel, we see the similar behaviors as those in Fig. \ref{fig:Ampq1}.
But now the amplification factor is much enhanced in the whole frequency
region by $q$. Even the higher $l$ modes can have significant amplification
factors. The superradiance region $0<\omega<\frac{qQ}{r_{+}}$ is
also enlarged by $q$. Interestingly, there appears a peak in the
amplification factor before it falls. When $Q=0.2$ and $\alpha=0.95$
(the nearly extremal black hole), this peak rises to nearly 100\%
above the amplification factor of the lower frequency. When $Q=0.8$,
the step-like behavior of the amplification factor becomes more obvious.
Note that though the superradiance region is changed by $\alpha$,
the amplification factor is almost unchanged with $\alpha$. All the modes satisfying
the condition (\ref{eq:SuperCondtion}) are amplified equally.

{\footnotesize{}}

\subsection{The condition for instability}

We have shown the superradiance of the 4D charged EGB black hole under
the charged scalar perturbation. Since the incident wave can be amplified
when its frequency satisfies (\ref{eq:SuperCondtion}), one may suspect
that the system is unstable. To clarify this, here we show the condition
for the   instability.

Multiplying (\ref{eq:radEq}) by the complex conjugate $\Psi_{0}^{\ast}$
on both sides and integrating it, we get
\begin{equation}
\left.\Psi_{0}^{\ast}\frac{\partial\Psi_{0}}{\partial r_{\ast}}\right|_{-\infty}^{\infty}+\intop_{-\infty}^{\infty}\left(\omega-\frac{qQ}{r}\right)^{2}\left|\Psi_{0}\right|^{2}dr_{\ast}=\intop_{-\infty}^{\infty}V_{\text{eff}}\left|\Psi_{0}\right|^{2}dr_{\ast}+\intop_{-\infty}^{\infty}\left|\frac{\partial\Psi_{0}}{\partial r_{\ast}}\right|^{2}dr_{\ast}.
\end{equation}
The right hand side is real. Using the boundary condition (\ref{eq:AmpBoundary})
and taking the imaginary part of both sides, we get a relation
\begin{equation}
(a^{2}+b^{2})^{1/4}\cos\left(\frac{1}{2}\arctan\frac{b}{a}\right)+\omega_{R}-\frac{qQ}{r_{+}}+\intop_{-\infty}^{\infty}2\omega_{I}\left(\omega_{R}-\frac{qQ}{r_{+}}\right)\Psi_{0}^{\ast}\Psi_{0}dr_{\ast}=0.
\end{equation}
 Here $\omega=\omega_{R}+i\omega_{I}$ and $a=\omega_{R}^{2}-\omega_{I}^{2}$,
$b=2\omega_{R}\omega_{I}$.
Since $\arctan\frac{b}{a}\in\left(-\frac{\pi}{2},\frac{\pi}{2}\right)$, the first term is always positive.

The instability implies $\omega_{I}>0$. When $\omega_{I}>0$, there must be $\omega_{R}<\frac{qQ}{r_{+}}$ to keep the above formula hold.  But when $\omega_{R}<\frac{qQ}{r_{+}}$, the sign of $\omega_{I}$ can not be determined. This means that the superradiance is the  necessary  but not sufficient condition for instability. To confirm
the   instability, one must ensure that the imaginary
part of the frequency is positive.

In the following section, we will show numerically that the eigenfrequencies
of the system have always negative imaginary part. Therefore, the
system is stable, though it has supperradiance. In fact, to trigger
the  instability, there should be an effective potential
well outside the event horizon to trap the reflected wave from
the near horizon region. However, it is easy to show that no potential
well outside the event horizon and thus no  instability of this background under charged scalar perturbations.

\section{The stability of the novel 4D charged EGB black hole}

We studied the amplification factor by considering the system as a
scattering process. To get the eigenfrequencies of the charged scalar
perturbation, we should take the system as an eigenvalue problem.
The boundary condition now was chosen as
\begin{align}
\Psi_{0}\to & \begin{cases}
e^{-i\left(\omega-\frac{qQ}{r_{+}}\right)r_{\ast}}\sim\left(r-r_{+}\right)^{-\frac{i}{2\kappa}\left(\omega-\frac{qQ}{r_{+}}\right)}, & r\to r_{+},\\
e^{i\omega r_{\ast}}\sim r^{i\omega r_{+}}e^{i\omega r}, & r\to\infty,
\end{cases}\label{eq:AIMasymp}
\end{align}
 where we used
\begin{equation}
r_{\ast}\to\begin{cases}
\frac{1}{2\kappa}\ln(r-r_{+}), & r\to r_{+}.\\
r+r_{+}\ln(r-r_{+}), & r\to\infty.
\end{cases}
\end{equation}
 There is only ingoing waves near the event horizon and outgoing waves
at the infinity. The system is dissipative and the frequency of the
perturbations will be the composition of quasinormal modes (QNMs).
Many numerical methods are developed to solve the quasinormal modes,
such as the shooting method, the WKB approximation method, the Horowitz-Hubeny
method and the continued fraction method (CFM) \cite{Konoplya2011}.
But here we will use the asymptotic iteration method (AIM) \cite{Cho2010}.

\subsection{The asymptotic iteration method}

The AIM was used to solve the eigenvalue problem of the homogeneous
second order differential equation \cite{Ciftci2003}.
Let us introduce a new variable first
\begin{equation}
\xi=1-\frac{r_{+}}{r}.
\end{equation}
 It ranges from 0 to 1 as $r$ runs from the event horizon to the
infinity. The radial equation (\ref{eq:radEq}) becomes
\begin{align}
0= & \frac{f(1-\xi)^{2}}{r_{+}}\left(\frac{\partial^{2}\Psi_{0}}{\partial\xi^{2}}\frac{f(1-\xi)^{2}}{r_{+}}+\frac{\partial\Psi_{0}}{\partial\xi}\frac{(1-\xi)^{2}\partial_{\xi}f-2f(1-\xi)}{r_{+}}\right)\nonumber \\
 & +\left(\left(\omega-\frac{qQ}{r_{+}}(1-\xi)\right)^{2}-\frac{(1-\xi)^{2}}{r_{+}^{2}}f\left(l(l+1)+(1-\xi)\partial_{\xi}f\right)\right)\Psi_{0}.
\end{align}
 The solution satisfying the asymptotic behavior (\ref{eq:AIMasymp})
in terms of $\xi$ has the following form
\begin{align}
\Psi_{0}= & \xi^{-\frac{i}{2\kappa}\left(\omega-\frac{qQ}{r_{+}}\right)}\left(\frac{1}{1-\xi}\right)^{i\omega r_{+}}e^{i\frac{\omega r_{+}}{1-\xi}}\chi(\xi),
\end{align}
 in which $\chi(\xi)$ is a regular function of $\xi$ in range $(0,1)$.
Function $\chi(\xi)$ obeys a homogeneous second order differential
equation
\begin{eqnarray}
\frac{\partial^{2}\chi}{\partial\xi^{2}} & = & \lambda_{0}(\xi)\frac{\partial\chi}{\partial\xi}+s_{0}(\xi)\chi,\label{eq:AIMeq}
\end{eqnarray}
 in which the coefficients
\begin{align}
-\lambda_{0}(\xi)= & -\frac{i}{\kappa\xi}\left(\omega-\frac{qQ}{r_{+}}\right)+\frac{f'(\xi)}{f(\xi)}+\frac{2\left(\xi-i\xi r_{+}\omega+2ir_{+}\omega-1\right)}{(\xi-1)^{2}},\\
-s_{0}(\xi)= & -\frac{1}{(\xi-1)^{4}\xi^{2}}\left(\frac{(\xi-1)^{2}}{2\kappa}\left(\xi^{2}\left(2r_{+}\omega+i\right)-4\xi r_{+}\omega-i\right)\left(\omega-\frac{qQ}{r_{+}}\right)\right)\\
 & -\frac{1}{(\xi-1)^{4}\xi^{2}}\left(\frac{(\xi-1)^{4}}{4\kappa^{2}}\left(\omega-\frac{qQ}{r_{+}}\right){}^{2}+\xi^{2}r_{+}\omega\left(r_{+}\omega(\xi-2)^{2}+i(\xi-1)^{2}\right)\right)\nonumber \\
 & +i\frac{f'(\xi)}{(\xi-1)^{2}\xi f(\xi)}\left(\xi\left(2r_{+}\omega+i\right)-\frac{(\xi-1)^{2}}{2\kappa}\left(\omega-\frac{qQ}{r_{+}}\right)-\xi^{2}\left(r_{+}\omega+i\right)\right)\nonumber \\
 & -\frac{l(l+1)\xi}{(\xi-1)^{2}\xi f(\xi)}+\frac{1}{(\xi-1)^{4}f(\xi)^{2}}\left((\xi-1)qQ+r_{+}\omega\right){}^{2}.\nonumber
\end{align}
 The coefficients $\lambda_{0}(\xi)$ and $s_{0}(\xi)$ are analytical
functions in the interval $(0,1)$. Now differentiating (\ref{eq:AIMeq})
with respect to $\xi$ iteratively, we get an $(n+2)$-th order differential
equation
\begin{equation}
\chi^{(n+2)}=\lambda_{n}(\xi)\chi'(\xi)+s_{n}(\xi)\chi(x),
\end{equation}
 where the coefficients can be determined iteratively.
\begin{align}
\lambda_{n}(\xi)= & \lambda'_{n-1}(\xi)+s_{n-1}(\xi)+\lambda_{0}(\xi)\lambda_{n-1}(\xi),\label{eq:AIMiter}\\
s_{n}(\xi)= & s'_{n-1}(\xi)+s_{0}(\xi)\lambda_{n-1}(\xi).\nonumber
\end{align}
 In asymptotic iteration method, the cutoff of the iteration for large enough
$n$ is determined by
\begin{equation}
\frac{s_{n}(\xi)}{\lambda_{n}(\xi)}=\frac{s_{n-1}(\xi)}{\lambda_{n-1}(\xi)}.
\end{equation}
 Then the quasinormal modes $\omega$ can be worked out from this
``quantization condition''. To improve the efficiency and accuracy,
let us making an expansion around a regular point $\xi=\xi_{0},$
\begin{equation}
\lambda_{n}(\xi)=\sum_{j=0}^{\infty}c_{n}^{j}(\xi-\xi_{0})^{j},s_{n}(\xi)=\sum_{j=0}^{\infty}d_{n}^{j}(\xi-\xi_{0})^{j}.
\end{equation}
The expansion coefficients $c_{n}^{j}$ and $d_{n}^{j}$ are functions
of frequency $\omega$. Substituting the expansions into (\ref{eq:AIMiter}),
we get iterative relations between these expansion coefficients.
\begin{align}
c_{n}^{j}= & (j+1)c_{n-1}^{j+1}+d_{n-1}^{j}+\sum_{k=0}^{j}c_{0}^{k}c_{n-1}^{j-k},\\
d_{n}^{j}= & (j+1)d_{n-1}^{i+1}+\sum_{k=0}^{j}d_{0}^{k}c_{n-1}^{j-k}.
\end{align}
 In terms of the expansion coefficients, the ``quantization condition''
becomes
\begin{equation}
d_{n}^{0}c_{n-1}^{0}=d_{n-1}^{0}c_{n}^{0}.
\end{equation}
 From this ``quantization condition'' , we can get the frequency
$\omega$ of the perturbation. The efficiency and accuracy of AIM
depends on the expansion point $\xi_{0}$. We will ensure the reliability
of the results by varying the expansion point $\xi_{0}$ and the iteration
times $n$.

\subsection{The eigenfrequency of perturbation}


 {\footnotesize{}}
\begin{figure}[h]
\begin{centering}
{\footnotesize{}}%
\begin{tabular}{cc}
{\footnotesize{}\includegraphics[scale=0.5]{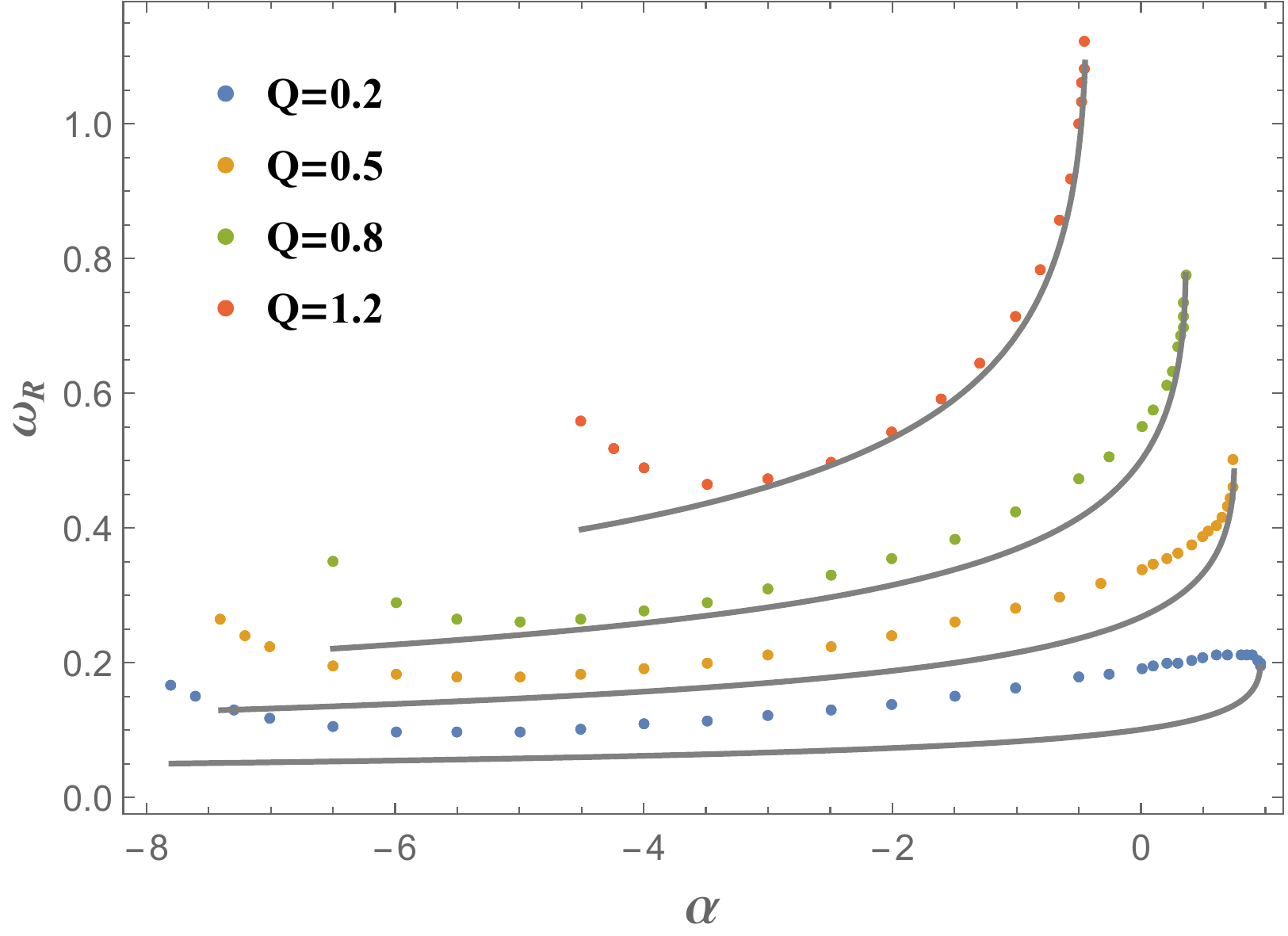}} & {\footnotesize{}\includegraphics[scale=0.5]{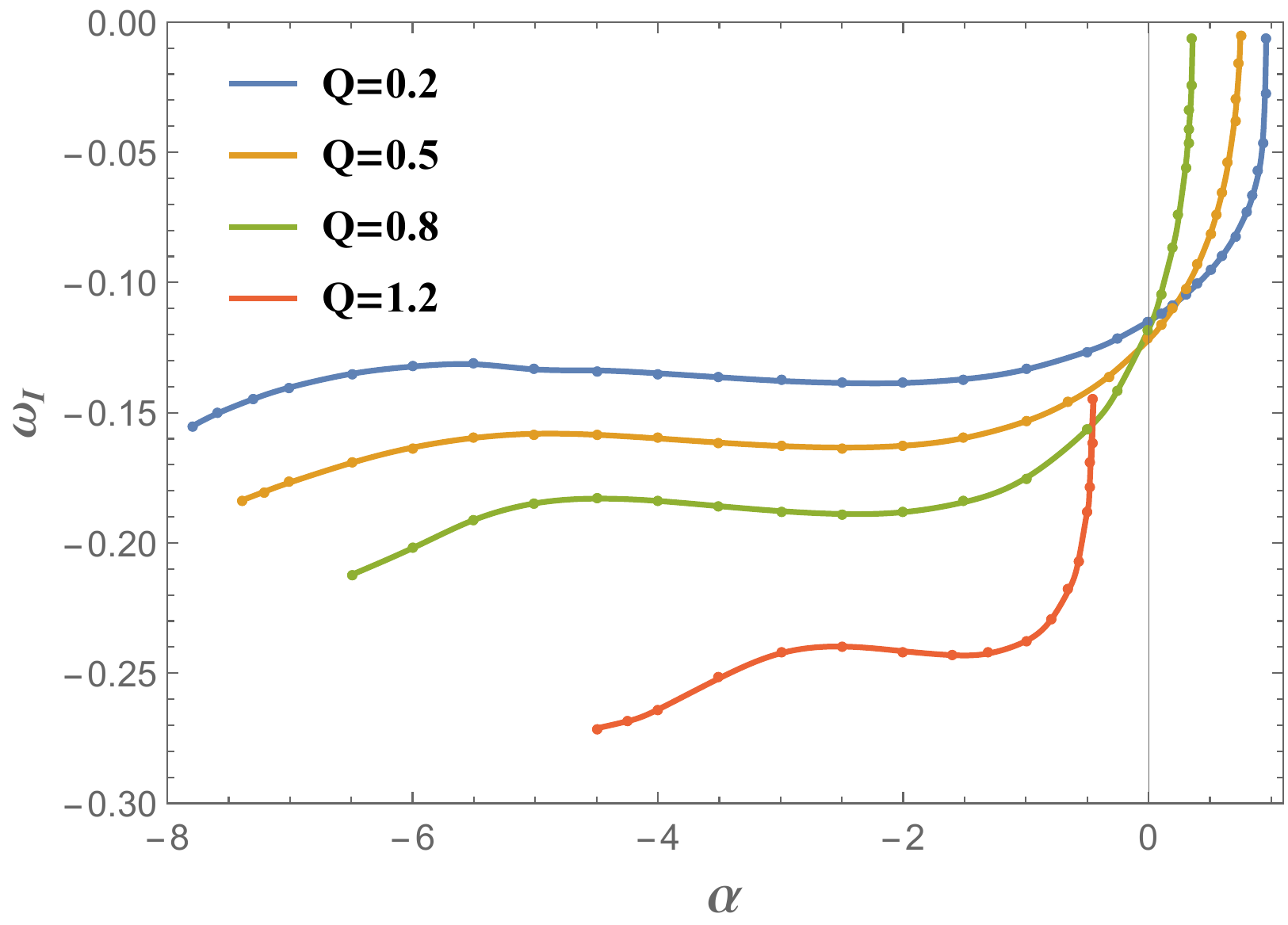}}\tabularnewline
\end{tabular}{\footnotesize\par}
\par\end{centering}
{\footnotesize{}\caption{\label{fig:wq1} Effects of $\alpha$ on the fundamental quasinormal
modes of the charged scalar perturbation when $q=1$. Left panel for
real part and right panel for imaginary part of the QNMs. We take
$Q=0.2,0.5,0.8$ and $1.2$ as examples to exhibit our results. When
$Q=0.2$, the range of $\alpha$ is $(-7.92,0.96)$. When $Q=0.5$,
the range of $\alpha$ is $(-7.49,0.75)$. When $Q=0.8$, the range
of $\alpha$ is $(-6.66,0.36)$. When $Q=1.2$, the range of $\alpha$
is $(-4.68,-0.44)$. We vary $\alpha$ in the reasonable regions,
respectively. The gray lines are corresponding threshold $\frac{qQ}{r_{+}}$
for superadiance since $r_{+}$ depends on $\alpha$.}
}{\footnotesize\par}
\end{figure}

We plot the fundamental quasinormal modes when $q=1$ in Fig. \ref{fig:wq1}.
In the left panel, we see that the real part $\omega_{R}$ of the
fundamental QNMs is a concave function of $\alpha$. It also increases
monotonically with $Q$. The most interesting point is that all $\omega_{R}$
live above the threshold $\frac{qQ}{r_{+}}$ of superradiance. For
positive $\alpha$ or large $Q$, the real part $\omega_{R}$ is very
close to the threshold. Therefore there should not be
instability of the charged 4D EGB black hole under perturbations.{\footnotesize{}
}We confirm this in the right panel by noticing that all the imaginary
part $\omega_{I}$ of the QNMs are negative. $\omega_{I}$ increases
with $\alpha$ at first and then keeps nearly unchanged before it
increases with $\alpha$ again. Interestingly, for positive $\alpha,$
$\omega_{I}$ tends to 0 rapidly when the black hole becomes extremal.
This implies the extremal 4D charged black hole may have normal modes.
For negative $\alpha$, the imaginary part $\omega_{I}$ can not reach
0 and rest on a finite negative value. These phenomenon also exists
for higher $l$.


{\footnotesize{}}
\begin{figure}[h]
\begin{centering}
{\footnotesize{}\includegraphics[scale=0.7]{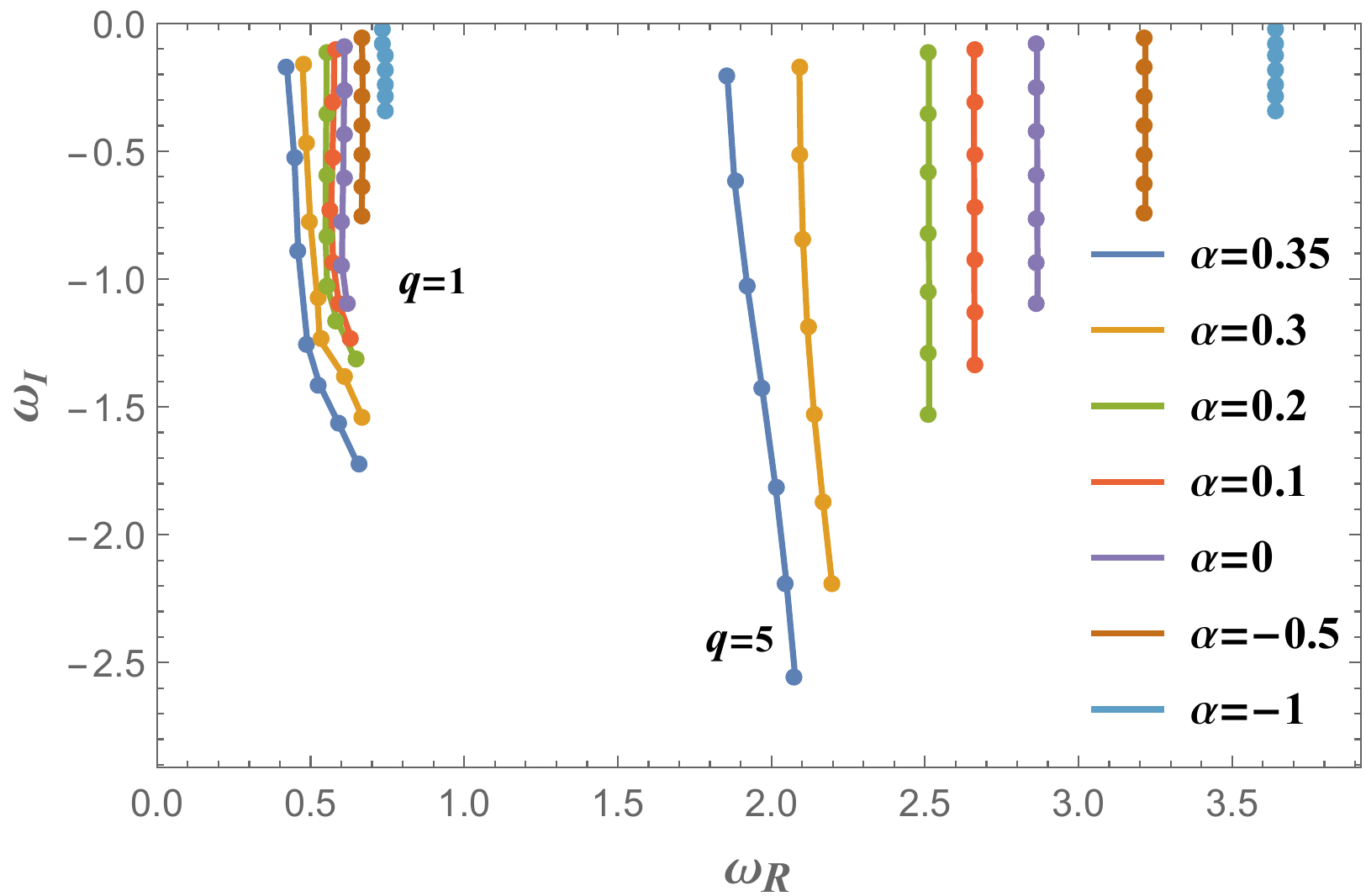}}{\footnotesize\par}
\par\end{centering}
{\footnotesize{}\caption{\label{fig:wOvertones} The overtones of the QNMs when $Q=0.8,l=0$.
$\alpha$ ranges from -1 to 0.35, where the black hole is nearly extremal.
The left hand side is for $q=1$, the right hand side for $q=5$.
The case of $\alpha=0$ corresponds to RN black hole.}
}{\footnotesize\par}
\end{figure}

We show the overtones of the QNMs when $Q=0.8$ and $l=0$ in Fig.
\ref{fig:wOvertones}. The green lines corresponds to the cases of
RN black hole. For positive $\alpha,$ we see that the the imaginary
part of the overtones increases with $\alpha$. Note that when $\alpha$
makes the black hole nearly extremal, the imaginary part of the  fundamental modes tends
to zero for both $q=1$ and $q=5$. For negative $\alpha$, the imaginary
part of the frequency decreases. When $q$ is small, the real part
of the overtones changes very small. While for large $q$, the real
part changes significantly.

\section{Summary }

We studied the parameter region of the novel 4D charged EGB black hole where allows the event horizon. The
black hole charge can be larger than the black hole mass due to the
existence of negative GB coupling constant $\alpha$.

With the help of appropriate boundary condition, the condition for superradiance was derived. We analyzed the effects of $\alpha$ on the amplification factor
of the superradiance in detail. The positive $\alpha$ enhances the superradiance and the negative $\alpha$ suppresses  it.
Almost all the frequencies satisfying the superradiance condition region are
amplified equally. Beyond this region, the reflected wave vanishes rapidly
and the amplification factor falls like a step-like function. We analyzed
this phenomenon from the viewpoint of effective potential.

To confirm whether the 4D charged EGB black hole has   instability, we  worked out the QNMs of the system using the asymptotic
iteration method. We found that all the real part of the QNMs live beyond the superradiance condition (\ref{eq:SuperCondtion}). The imaginary part of QNMs are negative. Therefore here is no
instability. We studied the effects of $\alpha$ on the QNMs in detail.
For $\alpha$  that makes  black hole extremal, we found that there may be normal
modes. Further detailed study on this phenomenon is required.

We further proved the condition for  instability of the
4D charged EGB black hole. Though there is superradiance, the system
is stable under perturbations due to the absence of an effective potential
well outside the event horizon to accumulate the energy and bounce
them back to the black hole again. The potential well can be induced by an artificial mirror outside the horizon or by a massive scalar. It has been shown that asymptotically-flat charged black holes are stable against massive charged scalar perturbations \cite{Hod2013}, but unstable for mirror boundary  \cite{Herdeiro:1111}. Since the potential well also appears in asymptotic dS or AdS spacetime, 
one can expect the superradiant  instability of the
4D charged EGB-(A)dS black holes \cite{Zhu:2014sya,Konoplya:2014lha}. In the forthcoming paper, we will study them in detail.

\section{Acknowledgments}

C.-Y. Zhang is supported by Natural Science Foundation of China under
Grant No. 11947067. S.-J. Zhang is supported by National Natural Science Foundation of China (Nos. 11605155 and
11675144). MG and PCL are supported by NSFC Grant No. 11947210.
MG is also funded by China National Postdoctoral Innovation Program
2019M660278.

\end{document}